\documentclass[11pt, oneside]{article}   	% use "amsart" instead of "article" for AMSLaTeX format
\usepackage{geometry}                		% See geometry.pdf to learn the layout options. There are lots.
\usepackage{graphicx}				% Use pdf, png, jpg, or eps§ with pdflatex; use eps in DVI mode
		
\usepackage{mathtools}
\usepackage{amssymb}
 \usepackage{color}
    \usepackage{amsmath}
   \usepackage{amssymb}
   \usepackage{amsthm}
\usepackage{cite}

\date{\empty}							% Activate to display a given date or no date

\begin{document}
%\maketitle
%\section{}
%\subsection{}
\noindent
\textsc{\Large Reflections of the  observer and  the observed in quantum gravity}

\vspace{21pt}
\noindent
\textsc{Dharam Vir Ahluwalia}

\vspace{42pt}

\noindent
E-mail: d.v.ahluwalia@iitg.ac.in
\vspace{5.25pt}

\noindent
Department of Physics\\
Indian Institute of Technology Guwahati\\
Assam 781 039, India \\ \\
and\\\\
Centre for the Studies of the Glass Bead Game\\
Chaugon, Bir\\
Himachal Pradesh 176 077

\vspace{84pt}
A broad brush impressionistic view of physics from the vantage point of she living on a nearby dark-planet Zimpok is presented so as to argue that the observed and the observer are reflected in quantum gravity through a universal mass shared by neurones and a unification scale of the high 
energy physics.

\vfill

\noindent
\textcolor{cyan}{\sc Essay written for the Gravity Research Foundation 2017 Awards for Essays on Gravitation.}

\vspace{11pt}\noindent
\textcolor{cyan}{Noted added 15 May 2017: Selected for Honorable Mention.}
\newpage

\begin{quote}
\texttt{``That which walks and that in which it walks spring from the same unifying principle.''}

\hfill{\texttt{Folklore on the planet Zimpok}}
\end{quote}
\vspace{21pt}

 The unifying principle of the Zimpokians on our tiny little planet, Earth, is known as the Lorentz algebra. ``That which walks'' is a metaphor for matter fields, and that ``in which it walks'' is what we call Minkowski spacetime. The archival records of the ancient ones show that what we call as the Heisenberg algebra 
 \begin{equation}
 \left[x,p_x\right] = i \hbar \label{eq:ha}
 \end{equation}
 was obtained by the Zimpokians by studying the vibrational absorption lines of CO, and a few other diatomic molecules of a nearby molecular cloud. How they came to such detailed knowledge, despite being in the dark sector, is open to speculation and to our own scientific progress.
  Zimpokians also had no Theory of Special Relativity, instead they considered all solutions~\textemdash~we call them representations~\textemdash~of the Lorentz algebra to have the same footing. Each solution of what we call Lorentz algebra to them was a space whose metric they deciphered by solving certain constraints. Two of these constraints, in our language read
 \begin{equation}
 \left[\eta, {\displaystyle\boldsymbol{\mathfrak{J}}}_i\right] = 0,\quad \left[\eta, {\displaystyle\boldsymbol{\mathfrak{K}}}_i\right] = 0,\qquad i=x,y,z\label{eq:constraints}
 \end{equation}
 where $\eta$ is `space metric' and ${\displaystyle\boldsymbol{\mathfrak{J}}}$ and ${\displaystyle\boldsymbol{\mathfrak{K}}}$ are the generators of the rotations and boosts, respectively. To these they added additional constraints. For us those correspond to additional restrictions on $\eta$ as imposed by various discrete symmetries. 

 They did not assume a metric, they derived it. For what we call Minkowski space they discovered, on solving the just mentioned constraints,  that $\eta$ has the form 
 \begin{equation}
 \exp[i\alpha]\times\mbox{diag}\{+1,-1,-1,-1\},\quad\alpha\in\Re.\label{eq:eta}
 \end{equation}
 While the multiplicative phase~$\exp[i\alpha]$ is generally set to~$\pm 1$ on Earth its significance becomes important while studying four-vector fields.
 They discovered that such a space allowed massive fields to carry Casimir invariants of 2 with a three fold degeneracy and 0 with no degeneracy.\footnote{On Earth these are called spin one and spin zero sectors, but Zimpokians prefer to be more precise as ${\displaystyle\boldsymbol{\mathfrak{J}}}^2$ is not a Casimir of the 
 four-vector space, nor of the Lorentz algebra. Because it becomes an accidental Casimir for specific solutions of the Lorentz algebra most Earthians are eternally confused about spin. On such matters Zimpokians keep their silence.} They introduced a field that carried both of these sectors, and for each sector they let $\alpha$ float.  They then calculated vacuum expectation value of the time ordered product of this field and its adjoint at two different spacetime points and adjusted $\alpha$ in a sector-dependent manner to ascertain unitarity at all energies.
 
 They then found that when they constructed a field that contained Casimir invariants of 6, 2, and 0 they could adjust counterparts of 
 $\alpha$ for each Casimir sector of the bigger nine dimensional space and could again secure unitarity
 at all energies. Their quantum gravity programme was thus developed along different lines than ours here on Earth. Their quantum description of gravity contained Higg's-like fields and these did not alter success of the theory in the local neighbourhood of their star. 
 
 For those on Earth who may wish to reproduce their calculations, one may note that for every space vector, $\xi$, they define  a dual space vector,  $\overline\xi=\xi^\dagger\eta$ and exploit every freedom of phase that is allowed in such definitions. As such they do away with introducing contravariant and covariant vectors, and have a unifying formalism for all spaces -- whether these be vector, spinorial, or any of the other infinite possibilities.

 The rotations and boosts for all spaces, they also knew, carried a general and very simple formal form. In our language these read
\begin{equation}
\exp\left[i{\displaystyle\boldsymbol{\mathfrak{J}}}\cdot{\displaystyle\boldsymbol{\theta}}\right],\quad
\exp\left[i{\displaystyle\boldsymbol{\mathfrak{K}}}\cdot{\displaystyle\boldsymbol{\varphi}}\right].
\end{equation}
 The
 ${\displaystyle\boldsymbol{\theta}}$ and ${\displaystyle\boldsymbol{\varphi}}$ denote the respective parameters for rotation and boost. By looking at an infinite dimensional representation of rotational sector of the Lorentz group they decoded that
 \begin{equation}
 p_x = \frac{\hbar}{i}\frac{\partial}{\partial{x}}\label{eq:px}
 \end{equation}
 and came to the conclusion that the kinematical description of physical reality required a mechanics that satisfied (\ref{eq:ha}). Their inference from observations and from their theoretical framework thus reinforced each other. Each required a quantum framework.
 
 They also noted that eigenfunctions of $p_x$, and in general of ${\displaystyle\boldsymbol{p}}$, had a spatial periodicity and it was measured by the wavelength (where we take notational liberty to make their results more accessible to Earthians)
 \begin{equation}
 \lambda_{\mathrm{dB}} =\frac{h}{p}\label{eq:db}
 \end{equation}
 To Zimpokians our Heisenberg algebra and the de Broglie wave particle duality had the same algebraic roots, roots they traced to the algebraic properties of space as regards rotational symmetry.
 
 The Zimpokians also constructed a counterpart of the Dirac field and, in our language, they understood it as arising from expansion coefficients of the parity operator residing in a direct sum of two massive Weyl representation spaces. This was a major triumph for them for they lived in a world that to us appears dark. Our world was only one fifth in cosmic energy content as theirs. It made our  gravitational imprints significantly weaker for their astronomers. Thus discovery of what we call the Dirac field was a major achievement for them.  It was their dark matter field. Their own world was luminous and this luminosity was governed by certain phases that varied over that in which they walked. Furthermore, that which walked, that is the matter fields of their world arose from the eigenspinors of the charge conjugation operator, as we call it. These fields were fermionic, and their mass dimension was one, and not $3/2$. With our world its only interaction was gravitational, and with the Higgs.

Zimpokians noted  that (\ref{eq:ha}) contained an element from the space in which they walked, that is $x$, and another from another space in which their counterpart of the Dirac field, or any of other fields, lived. By carrying momentum (say, $p_x$), equation  (\ref{eq:ha}) was unstable with respect to gravitational effects -- for when these momenta were measured, a collapse of the wave function, and thus of the associated energy momentum tensor occurred. And in fact it mattered whether one measured `$x$' first, or `$y$', or `$z$', and  `$p_x$' second, or `$p_y$,' or `$p_z$'~\textemdash~and vice versa. After careful deliberations on these matters they came to the conclusion that their understanding of that in which they walked, and that which walked, must undergo a profound change. A change which in our language is known as non-commutativity of spacetime. Knowing how intricately
(\ref{eq:ha}) and (\ref{eq:db}) are related, their elders came to the logical conclusion that both of these expressions must change, and change in such a manner that $\lambda$ saturates to a length scale associated with a mass 
roughly $10^{19}$ times the  lightest atom known in the universe which carried only about $5 \%$ of the total energy budget of the universe. This was the mass of their neurones, and they postulated correctly that mass of neurones in technology-creating biological beings must be this order. The beings who observe the universe, and the universe they observe, thus they deciphered are governed by the same mass scale: one biologic, other constructed from the fundamental constant that dictates the energy splitting of all diatomic molecules, that is $\hbar$ of (\ref{eq:ha}), the constant that relates dimensions of electric charge with mass (that is, $\sqrt{G}$), and the speed of what we call dark light,~$c$.

Having so deciphered, they saw that all objects of the universe, whether they be dust particles, or neurones, are governed by a mechanics in which $\lambda_\mathrm{dB}$ of (\ref{eq:db}) saturates to a wavelength around a universal invariant length.\footnote{\label{footnote}In a toy setting one such modification in our language reads
\[
\lambda_{\mathrm{dB}} \to \lambda =\frac{\overline{\lambda}_{\mathrm{P}}}{\tan^{-1} \left( {\overline{\lambda}_{\mathrm{P}}}/\lambda_{\mathrm{dB}}\right) } {\Bigg\{}
\begin{array}{ll}
\to \lambda_\mathrm{dB} & \mbox{for the low energy regime}\\
\to 4 \lambda_\mathrm{P} & \mbox{for the Planck realm}
\end{array}
\]
where $\lambda_P$ is the usual Planck length $\sqrt{\hbar G/c^3}$, and 
$\overline\lambda_\mathrm{P} \coloneqq 2\pi \lambda_\mathrm{P}$ is the Planck circumference.}
It is associated with a universal mass scale hovering around mass of the neurones in technology-creating biological beings or in the vicinity of Planck mass. They saw these observers and the universe they study as reflection of what we call quantum gravity. 
Or more precisely, in quantum gravity they saw the reflection of the laws that govern biology and cosmos.
Everything around them and us, from the lakes, to the clouds, to the flowers, to rain, was a manifestation reflected in quantum gravity. Adopting their view, the transition from the quantum to classical happens at what we call the Planck mass. Such an object could be an Avogadro number of cooper pairs in a superconducting quantum device and described by a `Planck mass' wave function, or a neurone, or a proverbial hep particle accelerated to Planck energies. 
 \vspace{42pt}

 \noindent
 \textbf{Credits.} The discussion surrounding equations  (\ref{eq:constraints}) and  (\ref{eq:eta}) is an adaptation from reference~\cite{Ahluwalia:2016rwl}. The discussion following equation~(\ref{eq:eta}) is based on voluminous calculations contained in~\cite{Ahluwalia:2017dva}. Part of the discussion surrounding equations~(\ref{eq:px}) and (\ref{eq:db}) and footnote~\ref{footnote} is based on references~\cite{Kempf:1994su,Ahluwalia:2000iw,Ahluwalia:1993dd,Ahluwalia:2005jn}.  That the result presented in footnote 1 may lead to an element of coherence in the early universe and in 
 biological systems was first mentioned in reference~\cite{Ahluwalia:2000iw}. In creating the story for this essay we benefited from consulting reference~\cite{Sami:2013ssa,vanDishoeck:2004fd} and M. Sami's talk at ``Workshop on Aspects of Gravity and Cosmology,'' organised to coincide with the 60th birthday celebrations of  T. Padmanabhan (IUCAA, Pune).
 
 The form of the essay is an experiment on how one may conjecture on scientific musings of other civilisations and how one may make one's technical work more accessible to the 
wider theoretical physics community.

%\bibliography{GRF2017.bib}

\begin{thebibliography}{1}

\bibitem{Ahluwalia:2016rwl}
D.~V. Ahluwalia, {\it {The theory of local mass dimension one fermions of spin
  one half}},  {\em Adv. in Applied Clifford Algebras} ({2017, in press}), doi:10.1007/s00006-017-0775-1
  [\href{http://arxiv.org/abs/1601.0318}{{\tt arXiv:1601.0318}}].

\bibitem{Ahluwalia:2017dva}
D.~V. Ahluwalia, S.~Horvath, C.-Y. Lee, and D.~Schritt, ``Unpublished notes,
  manuscript in preparation.'' {2012-2017}.

\bibitem{Kempf:1994su}
A.~Kempf, G.~Mangano, and R.~B. Mann, {\it {Hilbert space representation of the
  minimal length uncertainty relation}},  {\em Phys. Rev.} {\bf D52} (1995)
  1108--1118, [\href{http://arxiv.org/abs/hep-th/9412167}{{\tt
  hep-th/9412167}}].

\bibitem{Ahluwalia:2000iw}
D.~V. Ahluwalia, {\it {Wave particle duality at the Planck scale: Freezing of
  neutrino oscillations}},  {\em Phys. Lett.} {\bf A275} (2000) 31--35,
  [\href{http://arxiv.org/abs/gr-qc/0002005}{{\tt gr-qc/0002005}}].

\bibitem{Ahluwalia:1993dd}
D.~V. Ahluwalia, {\it {Quantum measurements, gravitation, and locality}},  {\em
  Phys. Lett.} {\bf B339} (1994) 301--303,
  [\href{http://arxiv.org/abs/gr-qc/9308007}{{\tt gr-qc/9308007}}].

\bibitem{Ahluwalia:2005jn}
D.~V. Ahluwalia, {\it {A Freely falling frame at the interface of gravitational
  and quantum realms}},  {\em Class. Quant. Grav.} {\bf 22} (2005) 1433--1450,
  [\href{http://arxiv.org/abs/hep-th/0503141}{{\tt hep-th/0503141}}].

\bibitem{Sami:2013ssa}
M.~Sami and R.~Myrzakulov, {\it {Late time cosmic acceleration: ABCD of dark
  energy and modified theories of gravity}},  {\em Int. J. Mod. Phys.} {\bf
  D25} (2016), no.~12 1630031, [\href{http://arxiv.org/abs/1309.4188}{{\tt
  arXiv:1309.4188}}].

\bibitem{vanDishoeck:2004fd}
E.~F. van Dishoeck, {\it {ISO spectroscopy of gas and dust: From molecular
  clouds to protoplanetary disks}},  {\em Ann. Rev. Astron. Astrophys.} {\bf
  42} (2004) 119--167, [\href{http://arxiv.org/abs/astro-ph/0403061}{{\tt
  astro-ph/0403061}}].

\end{thebibliography}
%\bibliographystyle{JHEP}

\providecommand{\href}[2]{#2}\begingroup\raggedright\endgroup

\end{document}